# Improving Electroencephalogram-Based Deception Detection in Concealed Information Test under Low Stimulus Heterogeneity

Suhye Kim, Jaehoon Cheon, Taehee Kim, Seok Chan Kim, and Chang-Hwan Im

*Abstract*—The concealed information test (CIT) is widely used for detecting deception in criminal investigations, primarily leveraging the P300 component of electroencephalogram (EEG) signals. However, the traditional bootstrapped amplitude difference (BAD) method struggles to accurately differentiate deceptive individuals from innocent ones when irrelevant stimuli carry familiarity or inherent meaning, thus limiting its practical applicability in real-world investigations. This study aimed to enhance the deception detection capability of the P300-based CIT, particularly under conditions of low stimulus heterogeneity. To closely simulate realistic investigative scenarios, we designed a realistic mock-crime setup in which participants were familiarized with all CIT stimuli except the target stimulus. EEG data acquired during CIT sessions were analyzed using the BAD method, machine learning algorithms, and deep learning (DL) methods (ShallowNet and EEGNet). Among these techniques, EEGNet demonstrated the highest deception detection accuracy at 86.67%, when employing our proposed data augmentation approach. Overall, DL methods could significantly improve the accuracy of deception detection under challenging conditions of low stimulus heterogeneity by effectively capturing subtle cognitive responses not accessible through handcrafted features. To the best of our knowledge, this is the first study that employed DL approaches for subject-independent deception classification using the CIT paradigm.

## I. INTRODUCTION

IN criminal investigations, guilty suspects often possess exclusive knowledge of a crime [1]. Consequently, they commonly employ self-regulatory strategies to minimize negative outcomes and avoid being identified as perpetrators. For example, suspects might deliberately omit specific details [2] or withhold critical crime-related information, irrespective of whether investigators have disclosed evidence [3]. Therefore, developing robust quantitative methods for evaluating suspect statements is essential to effectively address and counteract the deceptive strategies employed by suspects.

The most widely used methods for detecting deception through biomedical signals are the polygraph-based Comparison Question Technique (CQT) and the electroencephalogram (EEG)-based Concealed Information Test (CIT) [4]. The polygraph assesses physiological indicators of arousal, such as heart rate, blood pressure, respiration rate, perspiration, and skin conductivity. However, polygraph-based CQT has notable limitations, as evaluations can be compromised by confirmation bias stemming from an examinee's criminal history or pre-test interviews, and can be manipulated by examinees employing countermeasures, such as performing mental arithmetic or lightly biting their tongues during questioning [5]. Conversely, the EEG-based CIT offers a more quantitative and objective approach that relies on statistical methods for detecting deception [6].

The CIT is a psychophysiological method derived from Lykken's Guilty Knowledge Test, which initially utilizes galvanic skin response measurements [7], [8]. Currently, the P300-based CIT predominantly distinguishes deceptive individuals by analyzing differences in amplitude within the P300 component across stimuli [9]. The P300 is an event-related potential (ERP) component typically observed in EEG signals approximately 250–500 ms after stimulus presentation [10], [11]. A greater positive amplitude of the P300 waveform indicates higher stimulus relevance, reflecting enhanced cognitive processing of salient information [12].

The CIT employs three distinct types of stimulus: (1) the target stimulus (denoted by TR), an irrelevant item designated to elicit focused attention; (2) the irrelevant stimulus (denoted by IR), serving as a control item without relevance of crime scene; and (3) the probe stimulus (denoted by PR), which is related to the crime scene and recognizable only by guilty individuals [13]. Under this paradigm, guilty individuals typically show a significant difference in P300 amplitude between PR and IR, as PR elicits a stronger orienting response, and consequently, higher P300 amplitudes [14], [15]. Conversely, innocent individuals generally exhibit minimal differences in response to PR and IR [16].

The bootstrapped amplitude difference (BAD) method is

This work was supported by the grant of Supreme Prosecutors' Office of Korea. *(Corresponding author: Chang-Hwan Im).*

Suhye Kim is with the Department of Electronic Engineering, Hanyang University, Seoul, Republic of Korea (e-mail: ksuh5479@hanyang.ac.kr).

Jaehoon Cheon is with the Department of Electronic Engineering, Hanyang University, Seoul, Republic of Korea (e-mail: cjswogns000@hanyang.ac.kr).

Taehee Kim is with the Department of Electronic Engineering, Hanyang University, Seoul, Republic of Korea (e-mail: kth25532@hanyang.ac.kr).

Seok Chan Kim is with the Department of Forensic Science Investigation, Supreme Prosecutors' Office, Seoul, Republic of Korea (e-mail: skymr1@gmail.com).

Chang-Hwan Im is with the Department of Biomedical Engineering, Hanyang University, Seoul, Republic of Korea and the Department of Electronic Engineering, Hanyang University, Seoul, Republic of Korea (e-mail: ich@hanyang.ac.kr).

This study was approved by the Institutional Review Board of Hanyang University (HYUIRB-202408-006-1).

2most widely used to quantitatively compare differences in P300 amplitude to determine stimuli with salient meanings [17], [18]. This approach resamples data with replacement to generate a statistical distribution of PR and IR for a given individual, which enables the detection of significant differences in the P300 amplitude between stimuli. The detection rate of the P300-based BAD method generally improves when the amplitude difference between the PR and IR is large for guilty individuals.

In controlled laboratory environments, mock-crime scenarios typically involve ensuring that participants have no prior exposure to IR, thereby maximizing the differences in amplitude within the P300 component between PR and IR [19], [20]. Under carefully controlled conditions, the PR can be distinctly identified relative to the IR in guilty individuals. Consequently, laboratory studies have often reported a high accuracy in distinguishing guilty suspects from innocent participants [21], [22]. However, achieving comparably robust differences in P300 amplitude between the PR and IR in actual investigative contexts is notably more difficult, even among guilty suspects. In real-world scenarios, certain IR items may carry inherent personal significance, and their mere recognition can trigger orienting responses [23], [14]. This reduction in stimulus heterogeneity between the PR and IR can negatively impact the accuracy of statistical BAD method based on P300 amplitude. Furthermore, the amount and type of information provided to suspects before conducting the CIT significantly influence the validity of the outcomes when employing the BAD method [24]. Consequently, the accuracy of deception detection may deteriorate when one or more IR possess familiarity or meaningful relevance to a suspect. These observations highlight inherent limitations of the traditional BAD method based on P300 amplitude in real-world investigative scenarios, particularly when stimulus heterogeneity cannot be assured. Thus, developing new approaches capable of reliable deception detection is imperative, even under conditions characterized by low stimulus heterogeneity between PR and IR.

Recent research has explored the use of machine learning (ML) and deep learning (DL) techniques for EEG-based deception detection [25], [26], [27]. ML methods typically depend on handcrafted features designed to capture distinctive data patterns, whereas DL methods automatically extract intricate features directly from the raw EEG signals. Both ML and DL methods could identify features that conventional BAD methods may struggle to detect. Despite increasing interest in ML and DL applications, relatively few studies have demonstrated successful deception classification, specifically within the EEG-based CIT paradigm [28]. Only two previous studies have applied DL to the EEG-based CIT [25], [26]. However, one study suffered from issues with data leakage by utilizing data from the same participant in both the training and testing phases, whereas the other employed a subject-dependent approach, which has limited generalizability. To the best of our knowledge, no comprehensive subject-independent DL studies have been reported to date. Furthermore, the effectiveness of ML and DL models in distinguishing guilty from innocent individuals under conditions of low stimulus heterogeneity remains largely unexplored.

In this study, we introduced a novel mock crime experimental scenario specifically designed to emulate real-world investigative conditions by creating low stimulus heterogeneity between IR and PR. We hypothesized that under this low-heterogeneity condition, the BAD method would yield lower classification accuracy among guilty participants. Additionally, we anticipated that both the ML and DL approaches would outperform the BAD method, given their capacity to leverage more complex and subtle features from EEG signals. To empirically validate our hypothesis, we conducted a refined mock crime paradigm involving 67 participants. The accuracy for classifying deception detection was subsequently compared between the BAD, ML, and DL methods, with DL further incorporating a novel strategy for data augmentation.

## II. METHODS

### A. Experiments

The experiment consisted of two steps: a mock crime mission and a CIT. Before the experiment, participants completed the Symptom Checklist-90-Revised (SCL-90-R), a 90-item self-report inventory that assesses psychological symptoms and distress across nine principal dimensions [29], [30], [31]. The mock crime mission lasted approximately 30 min, and the CIT took 90 min. All stages were conducted on the same day in separate rooms, and different operators administered the mock crime mission and the CIT.

*1) Participants*: A total of 67 healthy adults aged 20 to 29 years (34 men; mean age 23.60 ± 2.30 years) participated in the experiment. All participants had normal or corrected-to-normal vision to ensure accurate responses to on-screen stimuli. None of the patients exhibited any psychological symptoms of mental disorders, as confirmed by the SCL-90-R [31]. To control for potential cerebral asymmetries and account for the low prevalence of left-handed individuals in South Korea, only right-handed participants were recruited [32], [33]. All participants provided informed consent and received monetary reimbursement for their participation.

*2) Mock crime mission*: The mock crime scenario comprised three distinct stages: (1) mission selection, (2) mission attendance, and (3) evaluation of mission performance. To enhance the participants' immersion, an assistant was posed as an additional participant, creating the impression that two individuals were simultaneously engaging in the scenario.

Before the mission selection, participants were informed that there were four envelopes, two containing instructions for 'innocent' missions and two for 'guilty' missions. Participants assigned to the guilty group performed a simulated criminal activity, whereas all other scenario elements remained consistent across the groups.

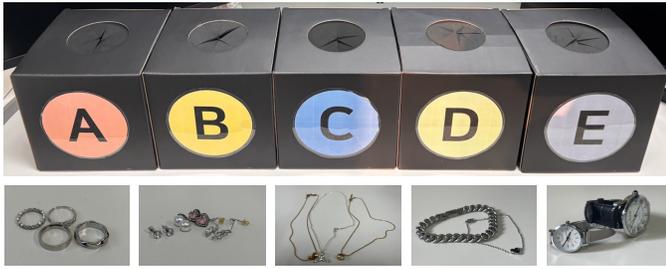

**Fig. 1.** The mystery boxes that participants were instructed to put their hands on and guess valuables (top) and the valuables that participants were shown through the testing paper (bottom).

In stage 1, each participant selected an envelope from a shuffled set and placed it in a pouch provided by the operator. This procedure prevented participants from discerning each other's assigned missions, although each envelope corresponded to a predetermined group assignment (i.e., innocent or guilty). Participants subsequently opened their envelopes during the phase of mission attendance. Both mission selection and attendance were performed in Office 1, with areas partitioned by a tall bookshelf to prevent participants from seeing the attendance area and mission-specific setups prematurely.

In stage 2, the participants individually engaged in the mission on one side of a partitioned area within Office 1, while the assistant waited separately in Office 2, which was a designated location for subsequent evaluation of mission performance. Upon opening their envelopes, the participants reviewed instructions directing them to reach into five "mystery boxes" and identify valuables inside by touch alone (Fig. 1). This process ensured that the participants encountered both probes and irrelevant items intended for subsequent CIT experiments. The five categories of variables are ring, bracelet, necklace, earring, and watch. Each mystery box contained at least three items in a single category, each with minor design differences (e.g., rings featuring varying jewel presence). The sequence of boxes was counterbalanced across participants to control for order effects. For instance, the order from boxes (A) to (E) varied among participants, with some encountering earrings, necklaces, rings, bracelets, and watches, and others experiencing a different sequence

Participants assigned to the guilty group were explicitly instructed to covertly steal a valuable from specified box (C) carefully avoiding detection by an ostensibly active CCTV camera installed in Office 1. In reality, the camera was inactive and displayed only a red indicator light, with no actual recording taking place. The duration of mission was limited to seven minutes for all participants [34]. After completing their assigned tasks, the participants proceeded to Office 2 for the evaluation of mission performance.

In stage 3, the participants were seated on partitioned desks and completed a written evaluation to assess their mission performance. The evaluation comprised two pages: the first contained five short-answer questions, and the second consisted of five matching questions. The participants were instructed to complete the tests sequentially. The assistant simulated the completion of the mission and completed an identical evaluation form, reinforcing the authenticity of the experimental scenario.

Subsequently, all the participants returned to Office 1, where they had originally been informed of the mission. The operator then opened and inspected the mystery boxes used in the mission. Following the inspection, the participants were informed that one of the variables was missing. Only those who answered all the second-page questions correctly and either declared innocence or remained silent were allowed to proceed to the next page. All participants were informed that their monetary reimbursement could be reduced from KRW 100,000 (approximately USD 75) to KRW 60,000 (approximately USD 45) if they were judged guilty in the upcoming test (i.e., CIT). Although only those who agreed to this change were allowed to continue, in practice, everyone who completed the CIT received the full KRW 100,000 regardless of the outcome.

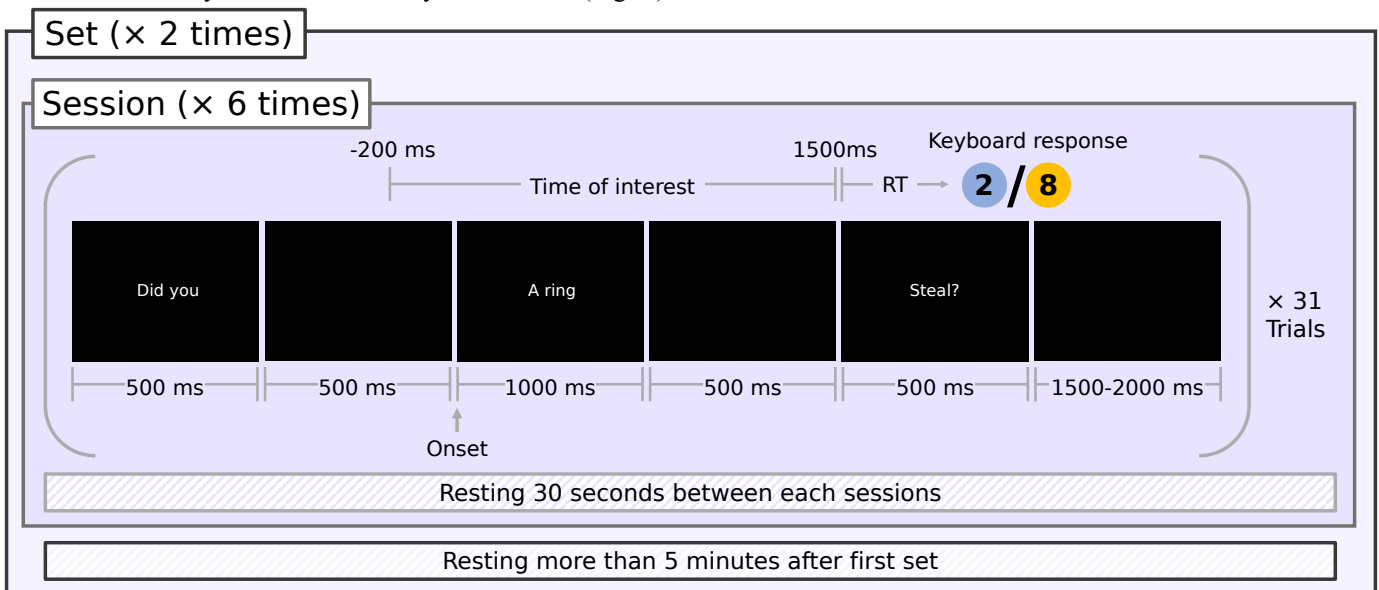

**Fig. 2.** The CIT procedure. The order of words follows Korean grammar (subject–object–verb). The stimulus "A ring" was replaced by one of six words: "A banknote," "A ring," "A bracelet," "A necklace," "Earrings," or "A watch."

*3) CIT*: Following the mock-crime mission, the participants moved to Office 3 for the CIT. The CIT consisted of three stages: pre-interview, CIT, and debriefing. The interview and testing stations were separated into partitions. Semi-structured pre-interviews were conducted. Once the participants declared their innocence, the operator explained the CIT procedure. After a brief practice trial, the CIT was administered, with EEG recording. This study employed a word-based CIT [35] rather than a picture-based approach, which allows its use even when direct physical evidence is unavailable.

Three phrases, "Did you," "Steal?," and "XXX," were presented in sequence on a monitor during the test. Because Korean language follows a subject–object–verb order, "XXX" appeared between "Did you" and "Steal?". "XXX" referred to "Banknote" or one of the five valuables used in the mock crime. For example, participants might see "Did you," "Steal?," and "A watch." They were instructed to press "8" if "XXX" was "Banknote" and to press "2" otherwise. They were not asked to answer yes or no to any sentence.

In this protocol, "Banknote" served as the TR for all participants. For the guilty group, the stolen item was designated a PR. For the innocent group, the same item was considered IR because they were unaware of which item was missing. Among innocent participants, PR was assigned in a counterbalanced manner based on the valuables they encountered, and all other items served as IR. Each participant had previously seen the stimuli corresponding to the mission items, except for the TR ("Banknote"), which preserved low heterogeneity among the CIT stimuli.

The participants were required to press either the "2" or "8" key immediately after the word "Steal?" appeared. Each word was displayed for 500 ms, except "XXX," which remained onscreen for 1000 ms. A 500 ms interstimulus interval followed each phrase, and a random interval of 1500–2000 ms with a blank screen separated consecutive sentences (Fig. 2). A total of 372 sentences were used, organized into 31 trials per session, with six sessions per set, and two sets in total (Fig. 2). Each session ended with a 30-second break, and the first trial always presented a randomly selected IR. The remaining 30 trials presented six stimulus types (one TR, one PR, and four IR), five times each in random order. A resting period of >5 min separated the two sets. In the second set, the participants pressed "2" when "Banknotes" appeared at the end of the sentence, reversing the initial response rule. If a participant moved their body excessively, an extra session was conducted before concluding the CIT.

After the CIT, the participants were debriefed by the operator. They were informed of the study objectives and of any false information used to enhance their immersion in the mock-crime task. This included a false notice that monetary compensation depended on the CIT results, the actual status of the envelope and camera, and the assistant's role as a disguised participant. Only participants who consented to the use of their data after being informed of the revised experimental objectives were included in subsequent analyses.

*B. Signal Processing*

*1) EEG signal acquisition*: EEG signals were recorded during the CIT experiment using a commercial biosignal recording system (Grael 4 K PSG: EEG; Compumedics, Victoria, Australia) at a sampling rate of 1,024 Hz. Thirty Ag–AgCl electrodes mounted on a Quick Cap (Quik-Cap Neo Net; Compumedics Neuroscan, Victoria, Australia) were positioned on the scalp according to the international 10–20 system (Fp1, Fp2, F11, F7, F3, Fz, F4, F8, F12, FT11, FC3, FCz, FC4, FT12, T7, C3, Cz, C4, T8, CP3, CPz, CP4, P7, P3, Pz, P4, P8, O1, Oz, and O2). The impedance of each electrode was maintained under approximately 5 k$\Omega$ and did not exceed 10 k$\Omega$. Vertical and horizontal electrooculograms (EOGs) were simultaneously recorded from two integrated bipolar leads.

*2) Signal preprocessing*: Preprocessing was conducted using the EEGLAB toolbox [36] and consisted of referencing, filtering, segmentation, and semi-manual inspection guided by independent component analysis (ICA), as follows:

1. Re-referencing EEG signals was re-referenced using the average amplitudes recorded at the left (M1) and right (M2) mastoids.
2. Filtering: A sixth-order Butterworth band-stop filter with cutoff frequencies of 59–61 Hz was applied to remove the power-line noise. This was followed by a sixth-order Butterworth bandpass filter with a cutoff frequencies of 0.1–50 Hz.
3. Segmentation: The signals were segmented into epochs from −200 ms before stimulus onset to 1,500 ms after onset (Fig. 2, "Time of Interest"). The baseline was corrected by subtracting the mean amplitude of the pre-stimulus interval (−200 to onset) for each epoch. The first epoch of each session was excluded from subsequent analyses.
4. Artifact rejection: ICA was performed to identify the physiological artifacts originating from the EOG and EMG signals. Before removing these components, the signals were downsampled to 512 Hz. A single researcher consistently rejected these artifacts from all the data.

*3) Dataset*: For the BAD analysis, EEG data from the Pz electrode, which exhibits the highest P300 amplitude, were selected [23], [13]. EEG segments exceeding ±75 μV were excluded from analysis. To maintain consistency, the remaining segments were standardized across all participants to match them with the fewest valid segments. Consequently, 46 segments per stimulus type (PR, TR, and combined IR) were retained, satisfying the minimum ERP analysis requirements [37]. To identify these 46 segments per stimulus, the mean amplitude within a 300–800 ms window was first calculated for all available segments [38]. The amplitude of each segment was then compared with this mean, and the 46 segments displaying the smallest deviations were selected for subsequent BAD analysis. For the IR category, each of the four subcategories contributed 11 or 12 segments, for a total of 46 segments.

The P300 amplitude was calculated for statistical analysis and ML using the peak-to-peak method [39]. After each segment was smoothed with a 100-ms window and a 2-ms stride, the maximum amplitude was identified within the 350–800-ms latency window. Subsequently, the minimum amplitude from this point to 1,500 ms was determined. P300 amplitude was defined as the difference between the maximum and minimum values. Finally, the P300 amplitudes constituted the dataset for BAD.

For the ML analysis, an equal number of segments was required for all electrodes during the same period. Therefore, segments exceeding ±75 μV on any electrode were removed. Ultimately, 44 epochs from stimulus onset to 1000 ms were selected for each stimulus to extract features. Candidate features were derived from the time, frequency, and network domains. The following features were extracted from each of the 44 segments: P300 amplitude, N200 amplitude, ratio of power spectral density (rPSD), differential entropy (DE), global clustering coefficient (gCC), and small-worldness.

The method used to extract the P300 amplitude was the same as that used for the BAD analysis. The N200 amplitude, which reflects cognitive processing, was calculated using Gamer and Berti's method [40]. The procedure for calculating the N200 amplitude was identical to that used for the P300, except that it was measured as the amplitude difference between the baseline and minimum amplitudes within a 200–350 ms latency windo

All features aside from the ERP components (P300 and N200) were computed for six frequency bands with a Hamming window and a sixth-order Butterworth filter. The frequency bands were low alpha (8–10 Hz), high alpha (10–13 Hz), low beta (13–16 Hz), mid beta (16–20 Hz), high beta (20–30 Hz), and gamma (30–50 Hz). rPSD was calculated as the ratio of the power spectral density (PSD) of each band to the total PSD between 8 and 50 Hz. DE was estimated by taking the logarithm of the power in each band [41]. The gCC was determined by averaging the local clustering coefficients, which measure the triangular connections around each node based on the phase-locking value. Finally, small-worldness was computed as the ratio of the gCC to the characteristic path length, using the Brain Connectivity Toolbox [42].

Each feature was then averaged across its respective segments. These features were then computed for each of the three stimuli (TR, PR, and IR) across all electrodes, resulting in 1,296 candidate features as follows: three stimuli × [(two ERP components × 30 electrodes) + (two spectral features × six frequency bands × 30 electrodes) + (two network features × six frequency bands)]. The DL architecture simultaneously processed the data from all three stimuli. To align with the DL approach, the differences between each pair of stimuli (PR minus TR, PR minus IR, and TR minus IR) in the ML input

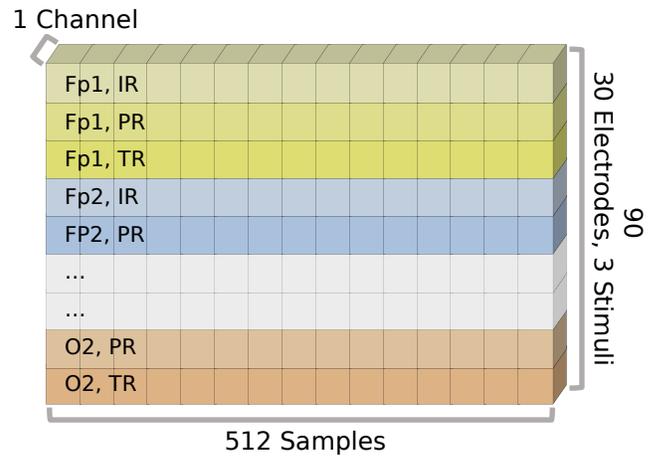

**Fig. 3.** Input data of DL.

dataset were calculated. This approach produced a 1,296-element feature vector.

The datasets for DL were identical before feature extraction. To prevent any single participant's data from dominating the DL model, we normalized each segment using the average and standard deviation computed from the same time segment for all electrodes. Next, the selected segments for each electrode were averaged for each participant to create ERP signals. A new strategy for data augmentation was introduced to reduce the potential overfitting caused by the limited size of the dataset. Specifically, 90% of the participants' 44 segments were combined with 10% of the segments randomly sampled from the other participants to expand the amount of training data. This method was repeated 45 times per participant to generate 45-fold more data. No segments from the participants assigned to the validation and testing sets were included in the data augmentation to avoid data leakage. The augmented data were then used to train the DL model. Finally, the training dataset comprised 28 participants × (1 original signal + 45 augmented signals) × 30 electrodes × 3 stimuli × 512 samples, and the test dataset comprised 2 participants × 1 original signal × 30 electrodes × 3 stimuli × 512 samples.

*C. Analysis and Evaluation*

*1) Statistical analysis*: The Wilcoxon rank-sum test was used to compare behavioral accuracy (bACC) between groups, and the Friedman test was used to examine differences in reaction time (RT) among the three stimulus types. We calculated bACC as the ratio of correct responses to the total number of trials for each stimulus. RT was measured from the appearance of the final word, "steal?" on the monitor for the participant's keyboard responses (Fig. 2). Trials with RT < 100 ms were excluded because such responses were considered predictive rather than physiological [43]. We also used the Wilcoxon rank-sum test to evaluate differences in P300 amplitude between groups, and the Friedman test was used to compare P300 and N200 amplitudes among the three stimulus types. When significant differences were observed, post hoc analyses were performed using the Wilcoxon signed-rank test. The *p* values were adjusted using the Bonferroni correction.

The BAD method was used to determine whether the P300

amplitude for PR exceeded that for IR for each participant [23]. A total of 46 trials for both PR and IR were resampled with replacement, and the trials for each stimulus were averaged to generate the ERP signals. Differences in P300 amplitudes between PR and IR were calculated using the respective ERP signals. This procedure was repeated 1,000 times to generate a bootstrap distribution. If the lower boundary of the 90% confidence interval for this difference was greater than zero, the participant was classified as guilty.

*2) ML*: Four ML architectures were employed: support vector machine (SVM), linear discriminant analysis (LDA), k-nearest neighbor (KNN), and random forest (RF). In this study, the KNN was configured to use five neighbors, and the RF was set to use 30 trees. From an initial set of 1,296 candidate features, we selected the top 20 features based on their Fisher scores [44]. Each model was trained and evaluated using subsets containing the top n features (n = 2, 3, …, 20), and the best performance was reported for each ML model.

*3) DL*: We used two widely-used convolutional neural network models, EEGNet and ShallowNet [45], [46], which have been extensively used in EEG-based classification problems. Minor modifications were made to adjust the models to fit the dataset. The input data for both models were structured as (B × C × E × T), where B is the batch size, C is the number of input channels, E is the number of electrode-stimulus features, and T is the sample length. In this study, each of the 30 EEG electrodes recorded signals for IR, PR, and TR, leading to 90 features for E. For example, electrode Fp1 contributed three features (Fp1–IR, Fp1–PR, and Fp1–TR), electrode Fp2 contributed another three (Fp2–IR, Fp2–PR, and Fp2–TR), and so on (Fig. 3). We then split the data into training, validation, and test sets at a 20:9:1 ratio.

The ShallowNet architecture consists of temporal convolution, spatial convolution, pooling, and classification layers. The temporal convolution layer applied 16 kernels (3, 384). Its kernel length was manually tuned to exceed half of the sampling rate and the duration of typical ERP components to ensure capture of the ERP signal. Unlike the original ShallowNet, which uses a one-dimensional temporal kernel (e.g., (1, 384)), we employed a two-dimensional kernel to integrate the signals from IR, PR, and TR. These three signals were then combined at the temporal convolution layer to enable the model to capture inter-stimulus relationships. The stride was set to (3, 1) to ensure that only one electrode was processed simultaneously. The spatial convolution layer contains 32 kernels. We replaced the original activation function with an exponential linear unit (ELU) function and added a batch normalization layer and ELU activation function after the two convolution layers [47]. The pooling layer uses average pooling with a window of 64 samples and a stride of two samples. ShallowNet was trained with a batch size of 32 and optimized using the AdamW optimizer at a learning rate of $5 \times 10^{-7}$.

The EEGNet model consists of three blocks. Block 1 includes a temporal convolution layer and a depth-wise convolution layer, block 2 contains a depth-wise convolution layer, and block 3 is the classification layer. In block 1, we employed a 2D temporal convolution layer with 32 kernels of size (3, 400). Similar to ShallowNet, we manually tuned the kernel length to exceed half the sampling rate and duration of typical ERP components. A kernel size of 3 was also introduced to capture the relationships among IR, PR, and TR. We set the stride to (3, 1) to ensure that only one electrode contributes to the learning process at a time. For depthwise convolution, we used a kernel size of (30, 1). We replaced the original batch normalization method in the EEGNet with group normalization by setting the number of groups to four [48]. In block 2, a depth-wise convolution layer with 32 kernels of size (1, 30) and a pointwise convolution layer with 32 kernels of size (1, 1) were employed. The four groups were normalized after the pointwise convolution layer. EEGNet was trained with a batch size of 16 and optimized using the AdamW optimizer at a learning rate of $1 \times 10^{-6}$.

All the hyperparameters described in this study were tuned using a random search. The dropout rates for both ShallowNet and EEGNet were set to 0.5. All the other layer compositions and hyperparameters, including the average pooling length and stride, were consistent with those of the original ShallowNet and EEGNet models. Training was continued for up to 300 epochs, and early stopping was triggered if the validation loss did not decrease by > 0.001 over 30 consecutive epochs. If the validation loss improved beyond this threshold, the training was continued until the maximum epoch was reached. All DL training and testing were implemented in Python 3.9 with PyTorch on an NVIDIA GeForce RTX 2080 Ti.

*4) Evaluation*: We assessed the classification performance of BAD, ML, and DL using accuracy, sensitivity, and specificity. For ML and DL, 30-fold cross-validation was used for evaluation, in which data from one guilty participant and one innocent participant were designated as the test set for each fold, whereas data from the remaining participants formed the training and validation set. The accuracy was calculated as the ratio of correctly classified participants to the total number of participants.

$$\text{Accuracy} = (TP + TN) / N, \quad (1)$$

where TP (true positives) represents the number of guilty participants correctly identified as guilty, TN (true negatives) denotes the number of innocent participants correctly identified as innocent, and N is the total number of participants (N = 60 in this study). Sensitivity was calculated using the following formula:

$$\text{Sensitivity} = TP / (TP + FN), \quad (2)$$

where FN (false negatives) is the number of guilty participants incorrectly classified as innocent. Specificity was calculated as follows:

$$\text{Specificity} = TN / (TN + FP), \quad (3)$$

where FP (false positives) is the number of innocent participants incorrectly classified as guilty.

## III. RESULTS

### A. Demographics and Psychological Measures

Seven participants were excluded because they failed to follow the operator's instructions (n = 6) or declined to continue
TABLE I

| DEMOGRAPHY, SCL-90-R SCORE OF EACH GROUP | | |
|---|---|---|
| | Innocent | Guilty |
| Sex(M/F) | 15 / 15 | 16 / 14 |
| Age(SD) | 23.6 (2.03) | 23.6 (2.76) |
| SCL-90-R(SD) | 41.65 (6.30) | 42.46 (5.34) |

the experiment (n = 1). All remaining participants achieved > 90% bACC on the CIT, implying that none were excluded based on CIT performance. The final analysis included data from 30 participants in the guilty group (15 men; mean age = 23.6 ± 2.03 years) and 30 participants in the innocent group (16 men; mean age = 23.6 ± 2.76 years). Detailed demographic information is presented in Table I. We examined the SCL-90-R scores to assess psychological symptoms. The score represents the average of T values across the nine principal symptom dimensions. No statistically significant demographic differences were observed between the innocent and guilty groups for any of the variables ($p > 0.1$).

### B. Statistical Analysis

*1) Behavioral data*: Table II presents the descriptive statistics for RT and bACC in both groups. No statistically significant differences in RT emerged between the guilty and innocent groups or among the three stimuli. Similarly, the bACC did not differ significantly by group or stimulus.

*2) Event-related Potential Component*: A Wilcoxon rank-sum test revealed no significant difference in the P300 amplitude between the guilty and innocent groups. However, Friedman test indicated significant differences among stimuli in both groups ($\chi^2(2) = 31.27$, $p < 0.0001$ in the guilty group; $\chi^2(2) = 39.27$ with $p < 0.0001$ in the innocent group). Post-hoc Wilcoxon signed-rank tests in the innocent group indicated that the P300 amplitude for TR exceeded those for the PR ($p < 0.0001$) and IR ($p < 0.0001$). In the guilty group, significant differences were observed between TR and IR ($p < 0.0001$), TR and PR ($p < 0.01$), and PR and IR ($p < 0.01$), with TR > PR > IR. Fig. 4 shows the grand-average ERP signals at the Pz and Cz electrodes for the three stimuli, and Fig. 5 shows the P300 amplitude at the Pz electrode for each stimulus along with its statistical significance.

A statistical analysis of N200 amplitude differences among stimuli at the Cz electrode was performed in a manner similar to the P300 analysis (bottom panel of Fig. 5). The Friedman test revealed significant differences in both the guilty group ($\chi^2(2) = 16.2$, $p < 0.001$) and the innocent group ($\chi^2(2) = 28.87$, $p < 0.0001$). Post-hoc Wilcoxon signed-rank tests showed that the

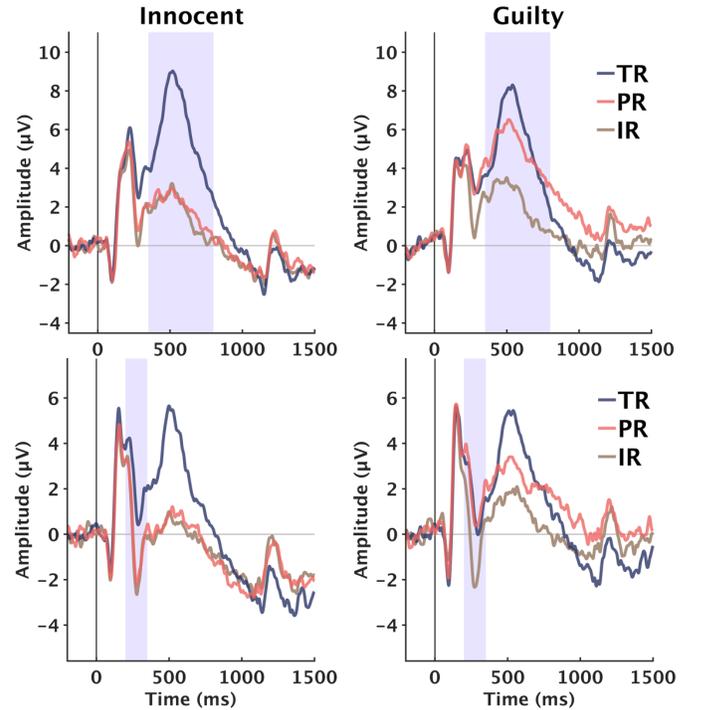

**Fig. 4.** Global Averaged ERP signals of Pz electrode (top) and Cz electrode (bottom) between innocent(left) and guilty (right) groups. The x-axis represents time in milliseconds (ms) and the y-axis represents amplitude in microvolts (μV). The black vertical line indicates stimulus onset. The purple shading denotes the interval used for calculating the highest peak of the P300 (Pz) and lowest peak of the N200 (Cz) amplitude, respectively.

P300 amplitude for TR was significantly lower than for IR in both groups ($p < 0.01$; innocent group, $p < 0.0001$ for the innocent group). Moreover, the guilty group exhibited a significantly lower P300 amplitude for PR than IR ($p < 0.01$).

*3) Bootstrapped amplitude difference*: The BAD method was applied in a participant-dependent manner to determine whether the P300 amplitude in PR exceeded that in IR. This approach yielded a classification accuracy of 73.5%, sensitivity of 50%, and specificity of 97%.

### B. Classification Analysis

Four ML models (SVM, LDA, KNN, and RF) and two DL models (ShallowNet and EEGNet) were evaluated in terms of cross-validation accuracy, sensitivity, and specificity. Table III summarizes these results, and Supplementary Tables S1 and S2 provide the classification details for individual participants across the BAD, ML, and DL approaches. Among the ML methods, LDA exhibits the highest accuracy (75.00%), followed by KNN (70.00%), SVM, and RF (both 68.33%). Among the DL architectures, ShallowNet exhibited an accuracy of 80.00%, whereas EEGNet achieved the best performance at 86.67%. Supplementary Fig. S1 presents the receiver operating characteristic curves for the BAD and ML methods.

TABLE II
REACTION TIME, BEHAVIORAL ACCURACY, P300, AND N200 FOR INNOCENT AND GUILTY GROUPS

| | Innocent | | | Guilty | | |
|---|---|---|---|---|---|---|
| | TR | PR | IR | TR | PR | IR |
| RT (ms) | 565.79 ± 328.85 | 568.02 ± 318.62 | 566.81 ± 318.94 | 692.04 ± 287.60 | 706.85 ± 280.24 | 711.09 ± 282.25 |
| bACC (%) | 99.56 ± 0.86 | 99.78 ± 0.56 | 99.93 ± 0.16 | 99.17 ± 1.11 | 99.67 ± 0.81 | 99.77 ± 0.23 |
| P300 (μV) | 12.30 ± 4.10 | 6.11 ± 2.15 | 6.11 ± 2.55 | 11.15 ± 4.45 | 8.15 ± 3.90 | 5.79 ± 2.47 |
| N200 (μV) | 0.13 ± 3.78 | 2.36 ± 3.45 | 2.64 ± 3.32 | 0.80 ± 4.39 | 0.81 ± 4.80 | 2.80 ± 3.81 |

## IV. Discussion

This study aimed to investigate classification models for effectively detecting deception in a realistic investigative scenario that showed robust performance under low-stimulus heterogeneity conditions in the CIT. The first was to design a mock crime scenario with low stimulus heterogeneity by having all participants perform the same variable-matching task. Under this condition, all participants were exposed to all test stimuli except the TR. A DL-based classification model, especially EEGNet, with a new strategy for data augmentation, could effectively classify deception. The EEGNet model correctly classified all the participants in the innocent group, except for one, thereby matching the performance of the BAD method. However, the EEGNet model achieved a significantly higher accuracy for participants in the guilty group than the traditional BAD method. Thus, a DL-based classification model with data augmentation could effectively identify deception using EEG-based CIT with low stimulus heterogeneity. This model could be applied in realistic investigative settings, even when investigators are uncertain whether the suspects possess prior knowledge or salient meaning regarding the test stimuli.

Statistical analysis of P300 amplitude revealed no significant difference in P300 amplitude for PR between the innocent and guilty groups (Fig. 5). This outcome may be attributed to our mock crime scenario, which reduced the heterogeneity among the PR and IR. Thus, in the guilty group, the PR may have been obstructed by the IR, preventing it from eliciting a sufficiently large P300 amplitude to distinguish it from the PR in the innocent group. These observations are consistent with prior findings, indicating that variations in the P300 response depend on the degree of stimulus heterogeneity [19], [20].

In the guilty group, the P300 amplitude of the PR exceeded that of the IR. Although both PR and IR were presented to all the participants, PR remained higher among guilty individuals, which is consistent with the findings of Farahani and Moradi [34]. Kim et al. have reported that the guilty group exhibited higher PR amplitudes than the IR in the P300 component, regardless of retroactive memory interference before the CIT [35]. Therefore, the orienting response to a stolen object in participants who intend to deceive could still be significantly more robust than the response to other IR, even under low heterogeneity between PR and IR [49], [15]. However, despite a statistically significant difference between the PR and IR amplitudes in the guilty group, the BAD method correctly identified only 50% of the guilty participants. Unlike group-level comparisons of amplitude differences, the BAD method classifies individuals in a participant-dependent manner, based on within-participant amplitude differences. Compared to group-level amplitude differences, low stimulus heterogeneity may have a stronger impact on participant-dependent analyses. These findings support our hypothesis that the BAD method may be insufficient for identifying suspects when stimulus heterogeneity is not ensured. By contrast, nearly all participants predicted to be innocent were classified correctly, in line with previous research showing that individuals without crime-related knowledge have high classification accuracy in BAD analyses [18].

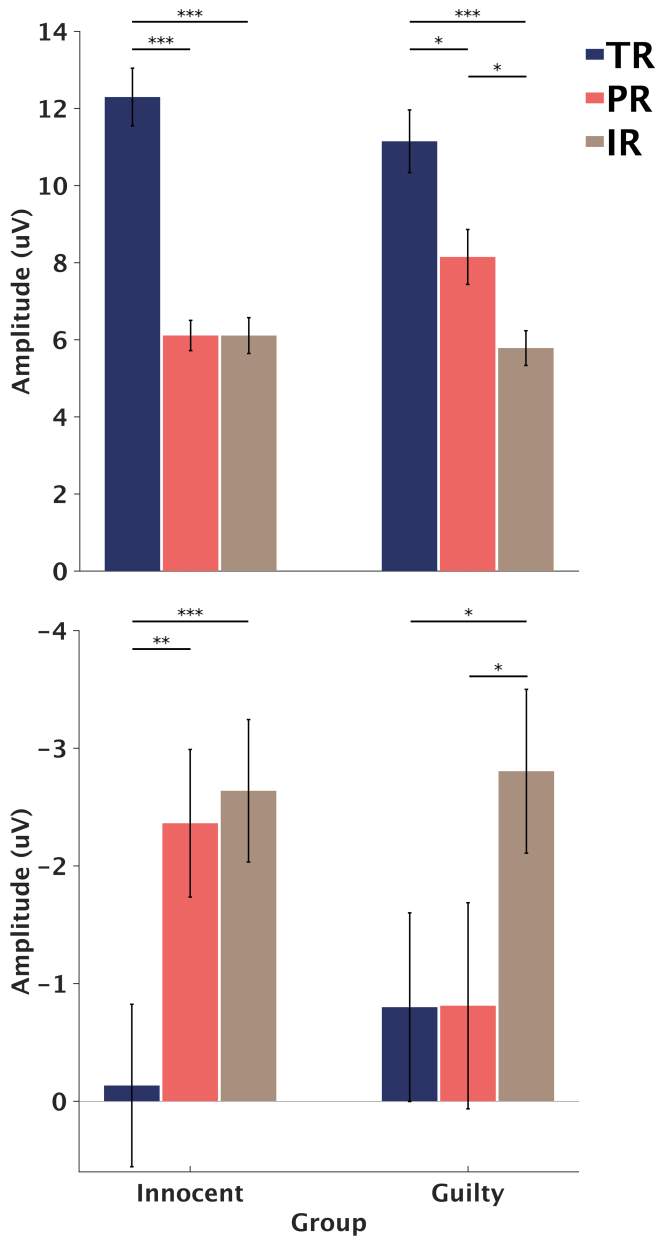

**Fig. 5.** The P300 amplitude at the Pz electrode (top) and the N200 amplitude at the Cz electrode (bottom) among stimuli in the innocent group (left) and the guilty group (right). *** $p < .0001$, ** $p < .001$, and * $p < .01$.

TABLE III
COMPARISON OF CLASSIFICATION RESULTS

| Model | Accuracy (%) | Sensitivity (%) | Specificity (%) |
|---|---|---|---|
| SVM | 66.67 | 63.33 | 70.00 |
| LDA | 75.00 | 73.33 | 76.67 |
| kNN | 70.00 | 60.00 | 80.00 |
| RF | 66.67 | 63.33 | 70.00 |
| ShallowNet | 80.00 | **80.00** | 80.00 |
| EEGNet | **86.67** | 76.67 | **97.00** |

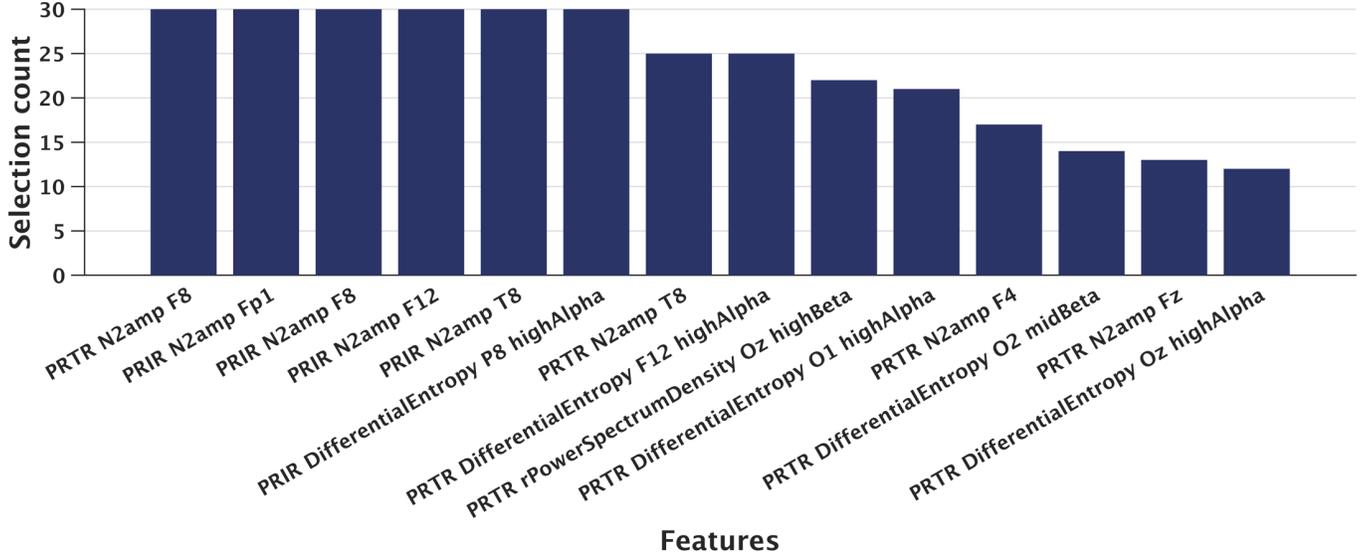

**Figure 6.** Number of times each feature was selected at the 30-fold cross-validation. PRTR: PR-TR, PRIR: PR-IR, N2amp: N200 amplitude, rPowerSpectrumDensity: the ratio of power spectrum density.

To address the limitations of the BAD method, we employed ML and DL approaches. The classification results indicated that the accuracy of LDA was comparable to that of the BAD method, although lower than previously reported results [27]. The selected features for model training (Fig. 6) showed that most were derived from the N200 component, which is another ERP component that is frequently examined in CIT research [50], [51]. Supplementary Fig. S2 presents the accuracy, sensitivity, and specificity for each ML method across the feature selections with respect to the number of features (n).

Based on the selected LDA features (Fig. 6) and the observed group differences in P300 and N200 amplitudes (Fig. 5), the LDA model likely used the amplitude difference between TR and IR to classify both guilty and innocent participants, and the amplitude difference between PR and IR to identify the guilty group. Although the ML performance was evaluated through a subject-independent analysis, and the BAD method relied on a within-participant approach, the ML accuracy was similar to that of the BAD method. Therefore, ML also required sufficient stimulus heterogeneity to achieve high classification accuracy.

Despite of the N200 component is also used for the CIT, whether differences between guilty and innocent groups in N200 amplitude are significant remains controversial [40], [50], [52]. Our results showed a statistically significant difference in the N200 amplitude, with opposite tendencies among stimuli being reported, compared with the P300 amplitude. Furthermore, the selected features for ML in our study were mainly derived from the N200 component. Thus, the N200 component may be used as an important feature for detecting deception in the CIT.

In this study, we employed the widely used EEGNet and ShallowNet models for EEG-based deception detection. Our EEGNet achieved an accuracy of 86.67% in distinguishing between guilty and innocent groups, surpassing the BAD method by 13.17%. This is the highest classification accuracy among the approaches analyzed in this study, even under subject-independent conditions. In this study, EEGNet outperformed ShallowNet, which accords with previous results in ERP analyses [46].

For the DL classification, we applied a new strategy for data augmentation to ERP signals to effectively enhance the performance of the DL model. We evaluated the performance of the DL model without applying it to assess the impact of data augmentation. In the absence of data augmentation, EEGNet and ShallowNet achieved the highest accuracies of 81.67% and 76.67%, respectively, which were lower than their performances with augmentation (Table IV). These results highlight the importance of data augmentation for improving DL performance. Given the potential drawbacks associated with lower deception detection accuracy in realistic investigational settings, employing augmentation techniques may be an effective strategy for boosting classification performance.

TABLE IV
PERFORMANCE OF DL MODELS WITHOUT AUGMENTATION

| Model | Accuracy (%) | Sensitivity (%) | Specificity (%) |
|---|---|---|---|
| ShallowNet | 76.67 | 70.00 | 83.34 |
| EEGNet | 81.67 | 76.67 | 86.67 |

The EEGNet model with data augmentation showed promise as an auxiliary tool for detecting subtle cognitive responses in examinees during real-world investigations, particularly in cases where identifying a suspect is challenging. Beyond suspect identification, the developed model could guide the investigative process by verifying whether a confessed suspect possesses specific knowledge of the physical evidence. This could be accomplished by selecting and presenting crime scene-related stimuli or items, such as weapons, thereby refining the investigation's direction.

Speed and accuracy are essential for the implementation of EEG-based deception detection in real-world scenarios. Constraints on space and time during the CIT may require data collection from fewer electrodes. In this study, the model was trained using the data from 30 electrodes. Therefore, developing a model with an acceptable performance and fewer electrodes would enable more convenient deception detection in real-world scenarios. In addition, providing clear explanations of classification decisions is vital for establishing the reliability and usability of DL in investigative contexts. Consequently, future work will explore explainable AI techniques to clarify classification decisions and verify their relationships with conventional ERP analyses [53], [54].

ACKNOWLEDGMENT

We thank Kang-Min Choi, Minsu Kim, Minyoung Chun, Wooseok Hyung, and Taegyeong Lee for their valuable feedback. We also thank Jiyoo Kim for her assistance with the experiments.

REFERENCES

[1] P. A. Granhag and M. Hartwig, "A new theoretical perspective on deception detection: On the psychology of instrumental mind-reading," *Psychol. Crime Law,* vol. 14, no. 3, pp. 189-200, 2008, doi: https://doi.org/10.1080/10683160701645181.

[2] M. Hartwig, *Interrogating to detect deception and truth: Effects of strategic use of evidence.* Psykologiska institutionen, Göteborgs universitet, 2005.

[3] M. Srivatsav, T. J. Luke, P. Granhag, L. Strömwall, and A. Vrij, "What to reveal and what to conceal? An empirical examination of guilty suspects' strategies," *PsyArXiv,* 2019, doi: https://doi.org/10.31234/osf.io/sx4nb.

[4] T. Brennen and S. Magnussen, "Lie detection: What works?," *Curr. Dir. Psychol. Sci.,* vol. 32, no. 5, pp. 395-401, 2023, doi: https://doi.org/10.1177/09637214231173095.

[5] W. G. Iacono and G. Ben-Shakhar, "Current status of forensic lie detection with the comparison question technique: An update of the 2003 National Academy of Sciences report on polygraph testing," *Law Hum. Behav.,* vol. 43, no. 1, p. 86, 2019, doi: https://doi.org/10.1037/lhb0000307.

[6] D. Wang, D. Miao, and G. Blohm, "A new method for EEG-based concealed information test," *IEEE Trans. Inf. Forensics Secur.,* vol. 8, no. 3, pp. 520-527, 2013, doi: 10.1109/TIFS.2013.2244884.

[7] D. T. Lykken, "The GSR in the detection of guilt," *J. Appl. Psychol.,* vol. 43, no. 6, p. 385, 1959, doi: https://doi.org/10.1037/h0046060.

[8] D. T. Lykken, "The validity of the guilty knowledge technique: The effects of faking," *J. Appl. Psychol.,* vol. 44, no. 4, p. 258, 1960, doi: https://doi.org/10.1037/h0044413.

[9] J. P. Rosenfeld, A. Ward, J. Wasserman, E. Sitar, E. Davydova, and E. Labkovsky, "Chapter 6 - Effects of Motivational Manipulations on the P300-Based Complex Trial Protocol for Concealed Information Detection," in *Detecting Concealed Information and Deception*, J. P. Rosenfeld Ed.: Academic Press, 2018, pp. 125-143.

[10] E. Donchin and M. G. Coles, "Is the P300 component a manifestation of context updating?," *Behav. Brain Sci.,* vol. 11, no. 3, pp. 357-374, 1988, doi: https://doi.org/10.1017/S0140525X00058027.

[11] J. Polich, "Updating P300: an integrative theory of P3a and P3b," *Clin. Neurophysiol.,* vol. 118, no. 10, pp. 2128-2148, 2007, doi: https://doi.org/10.1016/j.clinph.2007.04.019.

[12] R. Johnson Jr, "A triarchic model of P300 amplitude," *Psychophysiol.,* vol. 23, no. 4, 1986, doi: 10.1111/1469-8986.ep11022771.

[13] J. P. Rosenfeld, *P300 in detecting concealed information* (Memory detection: Theory and application of the Concealed Information Test). 2011, pp. 63-89.

[14] E. H. Meijer, F. T. Smulders, and A. Wolf, "The contribution of mere recognition to the P300 effect in a concealed information test," *Appl. Psychophysiol. Biofeedback,* vol. 34, pp. 221-226, 2009, doi: https://doi.org/10.1007/s10484-009-9099-9.

[15] I. Matsuda, H. Nittono, and T. Ogawa, "Identifying concealment-related responses in the concealed information test," *Psychophysiol.,* vol. 50, no. 7, pp. 617-626, 2013, doi: https://doi.org/10.1111/psyp.12046.

[16] L. A. Farwell and E. Donchin, "The truth will out: Interrogative polygraphy ("lie detection") with event-related brain potentials," *Psychophysiol.,* vol. 28, no. 5, pp. 531-547, 1991, doi: https://doi.org/10.1111/j.1469-8986.1991.tb01990.x.

[17] A. Bablani and D. Tripathi, *A review on methods applied on P300-based lie detectors* (Advances in Machine Learning and Data Science: Recent Achievements and Research Directives, no. 705). Berlin, Germany: Springer, 2018, pp. 251-257.

[18] J. P. Rosenfeld, "P300 in detecting concealed information and deception: A review," *Psychophysiol.,* vol. 57, no. 7, p. e13362, 2020, doi: https://doi.org/10.1111/psyp.13362.

[19] G. Ben-Shakhar, "Current research and potential applications of the concealed information test: an overview," *Front. Psychol.,* vol. 3, p. 342, 2012, doi: https://doi.org/10.3389/fpsyg.2012.00342.

[20] C. Wang *et al.*, "Memory detection with concurrent behavioral, autonomic, and neuroimaging measures in a mock crime," *Psychophysiol.,* vol. 61, no. 12, p. e14701, 2024, doi: https://doi.org/10.1111/psyp.14701.

[21] G. Ben-Shakhar and E. Elaad, "The validity of psychophysiological detection of information with the Guilty Knowledge Test: A meta-analytic review," *J. Appl. Psychol.,* vol. 88, no. 1, p. 131, 2003, doi: https://doi.org/10.1037/0021-9010.88.1.131.

[22] E. H. Meijer, N. K. Selle, L. Elber, and G. Ben-Shakhar, "Memory detection with the Concealed Information Test: A meta analysis of skin conductance, respiration, heart rate, and P300 data," *Psychophysiol.,* vol. 51, no. 9, pp. 879-904, 2014, doi: https://doi.org/10.1111/psyp.12239.

[23] J. P. Rosenfeld, M. Soskins, G. Bosh, and A. Ryan, "Simple, effective countermeasures to P300-based tests of detection of concealed information," *Psychophysiol.,* vol. 41, no. 2, pp. 205-219, 2004, doi: https://doi.org/10.1111/j.1469-8986.2004.00158.x.

[24] M. R. Winograd and J. P. Rosenfeld, "The impact of prior knowledge from participant instructions in a mock crime P300 Concealed Information Test," *Int. J. Psychophysiol.,* vol. 94, no. 3, pp. 473-481, 2014/12/01/ 2014, doi: https://doi.org/10.1016/j.ijpsycho.2014.08.002.

[25] A. Bablani, D. R. Edla, D. Tripathi, and V. Kuppili, "An efficient concealed information test: EEG feature extraction and ensemble classification for lie identification," *Mach. Vis. Appl.,* vol. 30, pp. 813-832, 2019, doi: https://doi.org/10.1007/s00138-018-0950-y.

[26] H. Wang *et al.*, "RCIT: An RSVP-based concealed information test framework using EEG signals," *IEEE Transactions on Cognitive and Developmental Systems,* vol. 14, no. 2, pp. 541-551, 2021, doi: https://doi.org/10.1109/tcds.2021.3053455.

[27] M. Žabčíková, Z. Koudelková, and R. Jašek, "Concealed information detection using EEG for lie recognition by ERP P300 in response to visual stimuli: A review," *WSEAS TIS&A,* 2022, doi: https://doi.org/10.37394/23209.2022.19.17.

[28] S. L. King and T. Neal, "Applications of AI-Enabled Deception Detection Using Video, Audio, and Physiological Data: A Systematic Review," *IEEE Access,* 2024, doi: https://doi.org/10.1109/access.2024.3462825.

[29] L. R. Derogatis and K. L. Savitz, "The SCL-90-R, Brief Symptom Inventory, and Matching Clinical Rating Scales," pp. 679–724, 1999.

[30] L. R. Derogatis and R. Unger, "Symptom checklist-90-revised," *The Corsini encyclopedia of psychology,* pp. 1-2, 2010, doi: https://doi.org/10.1002/9780470479216.corpsy0970.

[31] H. T. Won, J. H. Kim, K. J. Oh, C. T. Kim, Y. A. Kim, and M. Y. Kim, *Manual for the symptom checklist-90-revised (SCL-90-R).* Seoul, South Korea: Huno Co., Ltd., 2015.

[32] T. Guadalupe *et al.*, "Differences in cerebral cortical anatomy of left-and right-handers," *Front. Psychol.,* vol. 5, p. 261, 2014, doi: https://doi.org/10.3389/fpsyg.2014.00261.
[33] Neal. "How many people are left handed?" LeftyFretz. https://leftyfretz.com/how-many-people-are-left-handed/ (accessed Sep. 1, 2022).
[34] E. Farahani and M. Moradi, "A concealed information test with combination of ERP recording and autonomic measurements," *Neurophysiol.,* vol. 45, pp. 223-233, 2013, doi: https://doi.org/10.1007/s11062-013-9360-y.
[35] S. C. Kim *et al.*, "Retroactive memory interference reduces false positive outcomes of informed innocents in the P300-based concealed information test," *Int. J. Psychophysiol.,* vol. 173, pp. 9-19, 2022, doi: https://doi.org/10.1016/j.ijpsycho.2022.01.002.
[36] A. Delorme and S. Makeig, "EEGLAB: an open source toolbox for analysis of single-trial EEG dynamics including independent component analysis," *J. Neurosci. Methods,* vol. 134, no. 1, pp. 9-21, 2004, doi: https://doi.org/10.1016/j.jneumeth.2003.10.009.
[37] A. B. Dietrich, X. Hu, and J. P. Rosenfeld, "The effects of sweep numbers per average and protocol type on the accuracy of the P300-based concealed information test," *Appl. Psychophysiol. Biofeedback,* vol. 39, pp. 67-73, 2014, doi: https://doi.org/10.1007/s10484-014-9244-y.
[38] H. Kim, S. Y. Baik, Y. W. Kim, and S. H. Lee, "Improved cognitive function in patients with major depressive disorder after treatment with vortioxetine: A EEG study," *Neuropsychopharmacol. Rep.,* vol. 42, no. 1, pp. 21-31, 2022, doi: https://doi.org/10.1002/npr2.12220.
[39] M. Soskins, J. P. Rosenfeld, and T. Niendam, "Peak-to-peak measurement of P300 recorded at 0.3 Hz high pass filter settings in intraindividual diagnosis: complex vs. simple paradigms," *Int. J. Psychophysiol.,* vol. 40, no. 2, pp. 173-180, 2001, doi: https://doi.org/10.1016/S0167-8760(00)00154-9.
[40] M. Gamer and S. Berti, "Task relevance and recognition of concealed information have different influences on electrodermal activity and event-related brain potentials," *Psychophysiol.,* vol. 47, no. 2, pp. 355-364, 2010, doi: https://doi.org/10.1111/j.1469-8986.2009.00933.x.
[41] L. C. Shi, Y. Y. Jiao, and B. L. Lu, "Differential entropy feature for EEG-based vigilance estimation," in *2013 35th Annual International Conference of the IEEE Engineering in Medicine and Biology Society (EMBC)*, 3-7 July 2013 2013: EMBC, pp. 6627-6630, doi: 10.1109/EMBC.2013.6611075.
[42] M. Rubinov and O. Sporns, "Complex network measures of brain connectivity: uses and interpretations," *Neuroimage,* vol. 52, no. 3, pp. 1059-1069, 2010, doi: https://doi.org/10.1016/j.neuroimage.2009.10.003.
[43] R. Whelan, "Effective analysis of reaction time data," *The psychological record,* vol. 58, pp. 475-482, 2008, doi: https://doi.org/10.1007/BF03395630.
[44] A. R. Webb, *Statistical pattern recognition*. John Wiley & Sons, 2003.
[45] R. T. Schirrmeister *et al.*, "Deep learning with convolutional neural networks for EEG decoding and visualization," *Hum. Brain Mapp.,* vol. 38, no. 11, pp. 5391-5420, 2017, doi: https://doi.org/10.1002/hbm.23730.
[46] V. J. Lawhern, A. J. Solon, N. R. Waytowich, S. M. Gordon, C. P. Hung, and B. J. Lance, "EEGNet: a compact convolutional neural network for EEG-based brain–computer interfaces," *J. Neural Eng.,* vol. 15, no. 5, p. 056013, 2018, doi: 10.1088/1741-2552/aace8c.
[47] M. Kim and C.-H. Im, "HiRENet: Novel convolutional neural network architecture using Hilbert-transformed and raw electroencephalogram (EEG) for subject-independent emotion classification," *Comput. Biol. Med.,* vol. 178, p. 108788, 2024, doi: https://doi.org/10.1016/j.compbiomed.2024.108788.
[48] Y. Wu and K. He, "Group normalization," in *Proceedings of the European conference on computer vision (ECCV)*, 2018, pp. 3-19, doi: https://doi.org/10.48550/arXiv.1803.08494.
[49] G. Ben-Shakhar and E. Elaad, "The Guilty Knowledge Test (GKT) as an application of psychophysiology: Future prospects and obstacles," *Handbook of polygraph testing,* pp. 87-102, 2002.
[50] X. Hu, N. Pornpattananangkul, and J. P. Rosenfeld, "N 200 and P 300 as orthogonal and integrable indicators of distinct awareness and recognition processes in memory detection," *Psychophysiol.,* vol. 50, no. 5, pp. 454-464, 2013, doi: https://doi.org/10.1111/psyp.12018.
[51] J. P. Rosenfeld, X. Hu, E. Labkovsky, J. Meixner, and M. R. Winograd, "Review of recent studies and issues regarding the P300-based complex trial protocol for detection of concealed information," *Int. J. Psychophysiol.,* vol. 90, no. 2, pp. 118-134, 2013, doi: https://doi.org/10.1016/j.ijpsycho.2013.08.012.
[52] G. Ganis, D. Bridges, C.-W. Hsu, and H. E. Schendan, "Is anterior N2 enhancement a reliable electrophysiological index of concealed information?," *NeuroImage,* vol. 143, pp. 152-165, 2016, doi: https://doi.org/10.1016/j.neuroimage.2016.08.042.
[53] S. Bach, A. Binder, G. Montavon, F. Klauschen, K.-R. Müller, and W. Samek, "On pixel-wise explanations for non-linear classifier decisions by layer-wise relevance propagation," *PloS one,* vol. 10, no. 7, p. e0130140, 2015, doi: https://doi.org/10.1371/journal.pone.0130140.
[54] R. R. Selvaraju, M. Cogswell, A. Das, R. Vedantam, D. Parikh, and D. Batra, "Grad-CAM: visual explanations from deep networks via gradient-based localization," *Int. J. Comput. Vis.,* vol. 128, pp. 336-359, 2020, doi: https://doi.org/10.1007/s11263-019-01228-7.